# Broadband Reconfiguration of OptoMechanical Filters

**Parag B Deotare‡, Irfan Bulu‡, Ian W Frank‡, Qimin Quan‡, Yinan Zhang‡, Rob Ilic†, Marko Loncar‡**

‡ School of Engineering and Applied Sciences, Harvard University, Cambridge, MA 02138
† Cornell University, Ithaca, NY

**We demonstrate broad-band reconfiguration of coupled photonic crystal nanobeam cavities by using optical gradient force induced mechanical actuation. Propagating waveguide modes that exist over wide wavelength range are used to actuate the structures and in that way control the resonance of localized cavity mode. Using this all-optical approach, more than 18 linewidths of tuning range is demonstrated. Using on-chip temperature self-referencing method that we developed, we determined that 20 % of the total tuning was due to optomechanical reconfiguration and the rest due to thermo-optic effects. Independent control of mechanical and optical resonances of our structures, by means of optical stiffening, is also demonstrated.**

The combination of the advances in the fields of nanomechanics and nanophotonics has resulted in the recent emergence of the field of Nanoscale Optomechanical Systems (NOMS) [1-9], opening the door to revolutionary capabilities[1, 4-5, 10-11]. One such example, discussed in this letter, is a fully integrated, reconfigurable optical filter that can be programmed using internal optical forces. The fundamental building block of our platform is a doubly-clamped nanobeam mechanical resonator that is patterned with a one-dimensional lattice of holes to form high-quality factor optical nanocavity[12]. When two such resonators are placed in close proximity (Fig. 1) their optical modes couple resulting in sharp, wavelength-scale, optical resonances, which can be highly sensitive to the separation between the nanobeams[1, 13-14]. In addition, the structure supports propagating waveguide modes that can be excited over a wide wavelength range. These waveguide modes give rise to attractive (or repulsive) optical forces between the nanobeams, which in turn affects their mechanical configuration and results in the shift of the optical resonance of the filter [3, 7, 15-16]. In contrast to previous work, [5, 17-19] our waveguide-pump approach enables broadband operation in terms of actuation and tuning of the filter resonance.

Photonic crystal nanobeam cavities (PCNC) [12, 20-23] are well suited for the realization of optomechanical systems due to their small footprint, wavelength scale and high quality factor (Q) optical modes, small mass, and flexibility. These features allow for manipulation of optical signals as well as mechanical properties of PCNCs at low powers[14], a property which is of interest for dynamic signal filtering [24], routing, and modulation[25]. We designed the PCNC's using a deterministic approach [26], with a hole-to-hole spacing of 360 nm, a nanobeam width of 440 nm, and nanobeam separation of 70 nm. A SEM image of the final fabricated device is shown in Fig.1a with the inset showing the released cavity region (for fabrication details see Methods and Section 1 in supplementary material). This region supports propagating and localized modes with even (TE$_+$) and odd (TE$_-$) symmetry. Mode profiles of localized modes are shown in the inset of Fig. 1b. The same figure shows the dispersion of the even (red color) and odd (blue) mode as a function of the spacing between the nanobeams. It can be seen that the even mode is highly sensitive to the spacing while the odd mode is not [13]. Optomechanical coupling coefficients ($g_{om}$) were calculated to be 96 GHz/nm for the even mode, and $g_{om}$ of 0.73 GHz/nm for the odd mode (for 70nm separation). $g_{om}$ [1] is defined as the change in the resonance frequency due to a change in the separation of the coupled nano-beams. Fig. 1c shows a theoretical transmission spectrum of the device under excitation by the even electric-field guided mode, as well as the resulting optical force. One can see that localized cavity modes result in extremely large optical force due to the ultra-high quality factors associated with them [1]. This, however, comes at an expense of a very small operational bandwidth. Furthermore, reconfiguration of the filter results in shift of the high-Q cavity resonance, and wavelength-tracking mechanism is needed (Section 5 in supplementary material) to keep the pump laser in tune with the structure. This situation, analogous to common gain-bandwidth tradeoff issues in electronics (e.g. in amplifiers), can be overcome by taking advantage of broadband waveguide modes [3] supported at the long-wavelength range of our structure, where the optical force is approximately constant over a large wavelength range (Figure 1c). Our structure therefore combines unique advantages of both waveguide-based [3] and cavity-based [1, 5, 18] optomechanical systems, and has the following unique features: i) The wavelength of the *pump* beam can be chosen over wide wavelength range, so wavelength tracking is not needed to keep the *pump* in-tune with the structure during the reconfiguration process. Furthermore, broad-band light sources (such as LEDs) can be used as a *pump* and the same *pump* signal can be used to reconfigure a number of filters that operate at different wavelengths (possible applications in filter bank reconfiguration); ii) The wavelength scale, high-quality factor, localized modes have excellent sensitivity to mechanical motion, and tuning over several filter linewidths can be achieved using modest pump powers. Therefore operation at mechanical resonances (which also can require operation in vacuum) is not required [3]. (iii) Independent control of optical and mechanical degrees of freedom (optical and mechanical resonance) of the structure is possible; (iv) An inherent temperature self-referencing scheme can be used to individually evaluate the contributions from thermo-optical and optomechanical effects to the observed detuning in our filters. We do this by taking advantage of cavity modes with odd lateral symmetry which are *in-sensitive* to mechanical reconfiguration (Fig. 1b) [13-14], but have comparable thermo-optic sensitivities to the even cavity modes (see supplementary material).

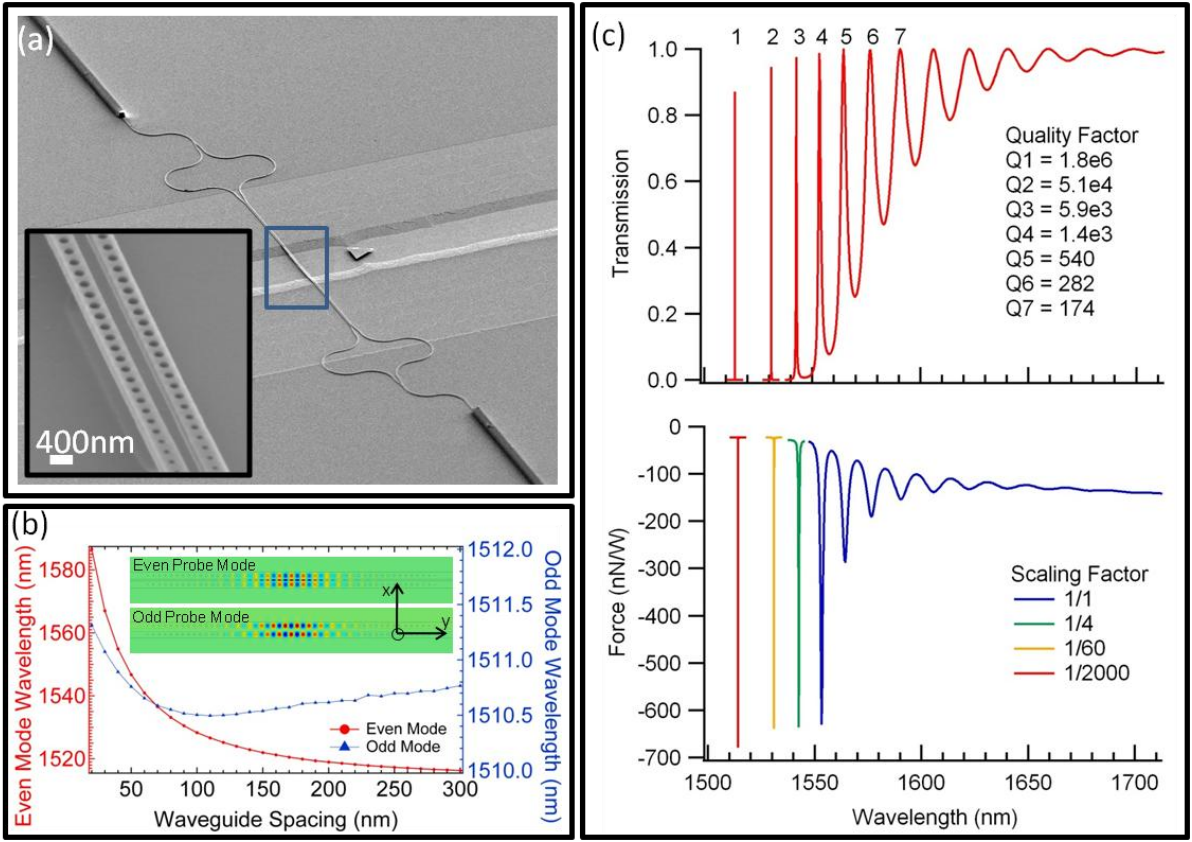

**Figure 1: Coupled Photonic crystal nanobeam cavities (a)** SEM image showing the complete device with the SU8 coupling pads, balanced Mach Zehnder Interferometer (MZI) arms, silicon waveguides and the suspended nanobeam cavity region. The inset shows the suspended nanocavity. **(b)** The red and blue curves show the dispersion of the even and odd cavity modes for various spacing between the two cavities. The even mode is highly dispersive while the odd mode is not. The device under test had a gap of 70 nm corresponding to an optomechanical coupling coefficient ($g_{om}$) of 96 GHz/nm for the even mode and 0.73 GHz/nm for the odd mode. The inset shows the profiles of dominant electric field (y) component of the two modes. The cavity modes are localized near the center of the nanobeams. **(c) top:** Simulated transmission of the device for the even electric-field profile. **bottom:** The corresponding optical force is in nN/W generated by the even mode for various pump wavelengths. The negative sign indicates the attractive nature of the force. The force for the first three modes has been rescaled (multiplied by the indicated scaling factor) for better comparison. The transmission spectrum and the mutual optical force between the nano-beams for low-Q modes ($Q < 10^4$) were calculated using finite-element simulations. The two highest-Q modes with quality factors $1.8 \times 10^6$ and $5.1 \times 10^4$ were treated using temporal coupled mode theory.

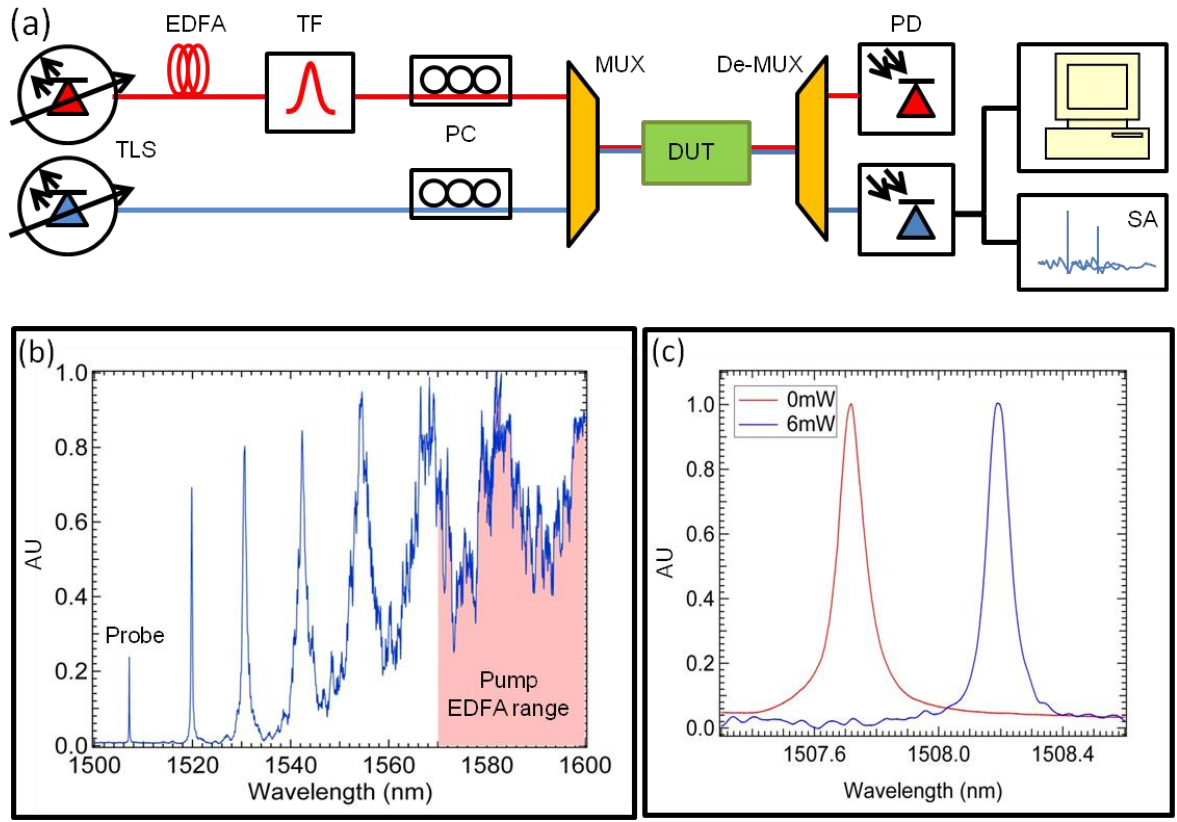

**Figure 2:** Experimental results for waveguide pump/cavity-probe system (a) Schematic of the pump-probe characterization setup used in the experiment. The cavity resonance was probed using a tunable laser (Santec TLS 510) and a fast, sensitive InGaAs detector (New Focus). Another tunable laser was used as the *pump* and was amplified using a high power output EDFA (L Band Manlight). The input and output WDMs (Micro Optics) ports were designed to operate in the 1470-1565 nm and 1570 -1680 nm ranges. The mechanical response of the devices was studied using a real time spectrum analyzer (Tektronix RSA 3300). TLS: tunable laser source, EDFA: Erbium doped fiber amplifier, TF: tunable filter, PC: polarization controller, MUX: multiplexer, De-MUX: de-multiplexer, PD: photo detector, DUT: device under test. (b) A transmission spectrum of the fabricated cavity. The fundamental even mode is found to be near 1507 nm with a Q of 15000. The shaded region shows the operating range of the EDFA used to amplify the *pump*. The *pump* has a waveguide like mode profile (See supplementary Section 4) since it is outside the bandgap and is used to induce mechanical deformation to detune the even mode cavity resonance. (c) Cavity resonance was tuned from 1507.72 nm to 1508.19 nm when 6 mW of *pump* power at 1583 nm was used. Cavity tuning was due to contribution from thermo-optical, nonlinear optical and optomechanical effects.

The devices were probed using a butt coupling technique using two lensed fibers. Unlike tapered fiber [27] probing, this type of characterization method remains invariant under any mechanical oscillations of the suspended waveguides and can be easily integrated for on-chip applications. Light from a tunable laser was amplified using an Erbium doped fiber amplifier (EDFA) and used as the *pump* while another tunable laser was used to *probe* the structure. A schematic of the experimental setup is shown in Fig. 2a.

A measured transmission spectrum of the device (similar to the one shown in Fig. 1a) is shown in Fig. 2b with the fundamental even mode at 1507.72 nm (The slight discrepancy in the measured and simulated resonance wavelength shown in Fig. 1c is due to the error associated in measuring the dimensions using a SEM image). The shaded portion under the transmission curve denotes the *pump* region which lies outside the bandgap of the photonic crystal. We note that due to the limited tuning range of the pump EDFA, in our experiments the *pump* is not positioned in the flat part of the transmission spectrum. Therefore, the *pump* benefits from the light confinement and low-Qs of extended resonances of the structure (broad peaks in the transmission spectrum). Based on theoretical predictions, we estimate that the force associated with this effect is approximately twice as big as the force in the "flat" part of the spectrum (Fig. 1c). When the *pump* laser excites only the even mode, an attractive optical force is generated which reduces the separation between nanobeams and red-detunes the even probe mode. Fig. 2c shows the experimental detuning of the fundamental even mode with a 6 mW *pump* (estimated power inside the waveguide just before the photonic crystal) at 1583 nm. This detuning is in part due to optomechanical (OM) effects, and in part due to the combination of thermo-optic (Th), free carrier dispersion (FCD) and Kerr effects. The overall detuning of the modes is given by:

$$\Delta_c^{'} = \Delta_c^{Th} + \Delta_c^{FCD} + \Delta_c^{Kerr} + \Delta_c^{OM} \qquad (1)$$

We theoretically estimated the contributions of these nonlinear effects and found that in addition to optomechanical detuning, thermo-optic effects play an important role in the detuning of the cavity resonance (both even and odd modes shift towards longer wavelengths with increasing temperature). For the given input power range, themo-optic effects can be as high as 80 % (refer to Section 2 in supplementary material). However, due to the nonlinear dependence of thermo-optic effect on pump power, its contribution increases at higher powers, while the optomechanical effect remains linear.

In order to decouple thermal and mechanical effects, we used an odd localized mode of the cavity as an on-chip temperature reference. This is possible since this mode's resonance does not tune with mechanical motion (Fig. 1b), and tunes only with temperature. The experimentally determined ratio of the temperature shifts of the even and odd mode is 0.93 (Section 3 in supplementary material), and is due to slightly

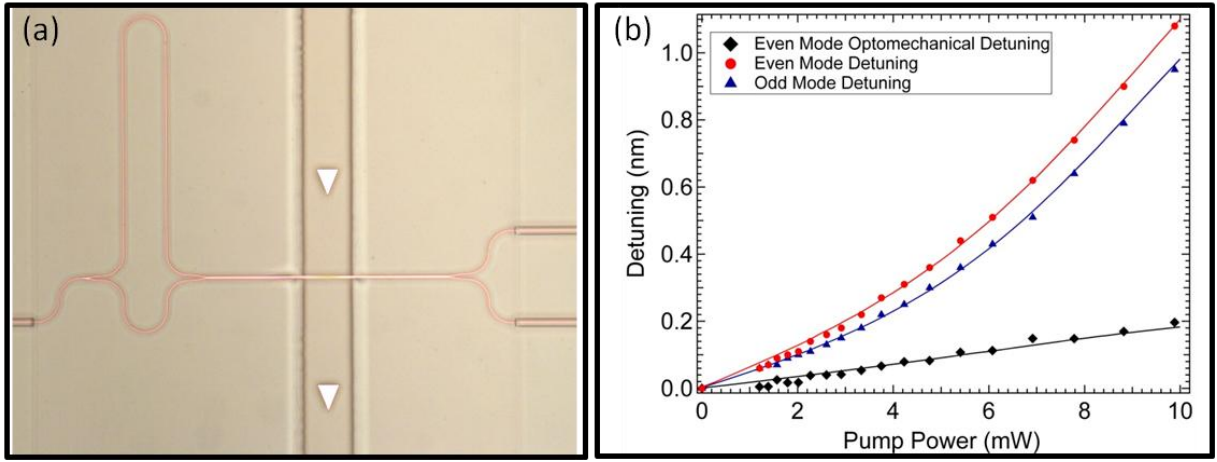

**Figure 3: Evaluation of optomechanical effect using temperature self-referencing:** (a) Optical micrograph of a device with an unbalanced MZI for simultaneous measurement of both the even and odd modes (b) Tuning of the cavity resonance for various *pump* powers and fixed *pump* wavelength of 1573.5 nm. The red circles and blue triangles correspond to even and odd modes while the black diamonds show the optomechanical detuning of the even mode. The latter was obtained by first determining the ratio of detuning for the even and odd modes due to purely thermal effects, by heating the sample (supplementary Section 3). Since the odd mode was sensitive only to temperature, it was used as a temperature sensor and the even mode temperature detuning was estimated using the above ratio and subtracted from the total detuning of the even mode. Using this self-referencing technique we estimated that more than 20 % of the total tuning is due to purely optomechanical effects

different overlaps of the two modes with the silicon nanobeams.

In order to simultaneously excite the even and the odd probe modes, and use the odd cavity mode as a built-in temperature reference, we incorporated an unbalanced Mach Zehnder Interferometer (MZI) before the cavities (Section 4 in supplementary material). This scheme is also useful to actuate the device using odd pump mode and generate repulsive force. However, the $g_{om}$ resulting from odd modes is extremely small and the repulsive force will cause very small, un-measurable, shifts in cavity resonance do to the optomechanics. Fig. 3a shows an optical micrograph of the new configuration. In this case the cavity resonance (even mode) was at 1524.9 nm. Fig. 3b shows the experimentally measured detuning of the *probe* modes (even and odd) for various *pump* powers (estimated power inside the waveguide just before the photonic crystal) and *pump* wavelength of 1573.5nm. The change in the resonances of both the even and odd modes are denoted with red circles and blue triangles respectively, while the shift resulting solely from the optomechanical effect is shown with black diamonds. The latter was obtained as proportional difference of the shifts of even and odd modes. The solid lines in Fig. 3b were obtained using Eqn. 1. Thermal, FCD, Kerr and optomechanical contributions of the *pump* and *probe* along with the cross coupling terms were included. The change in stored energy due to self-detuning of the *pump* was also computed and included in the estimation (Section 2 in supplementary material). The significance of thermo-optic effects is clearly seen, and 80% of tuning can be attributed to this effect. One way of reducing the relative contribution of thermo-optical effects in our devices is to improve $g_{om}$ by using nanobeams with smaller gaps. This however, is challenging with our current fabrication capabilities. Another method is to excite the structure with pulsed *pump* beam with repetition rates faster than the thermal response but lower than the mechanical resonance of the structure (Section 6 in supplementary material).

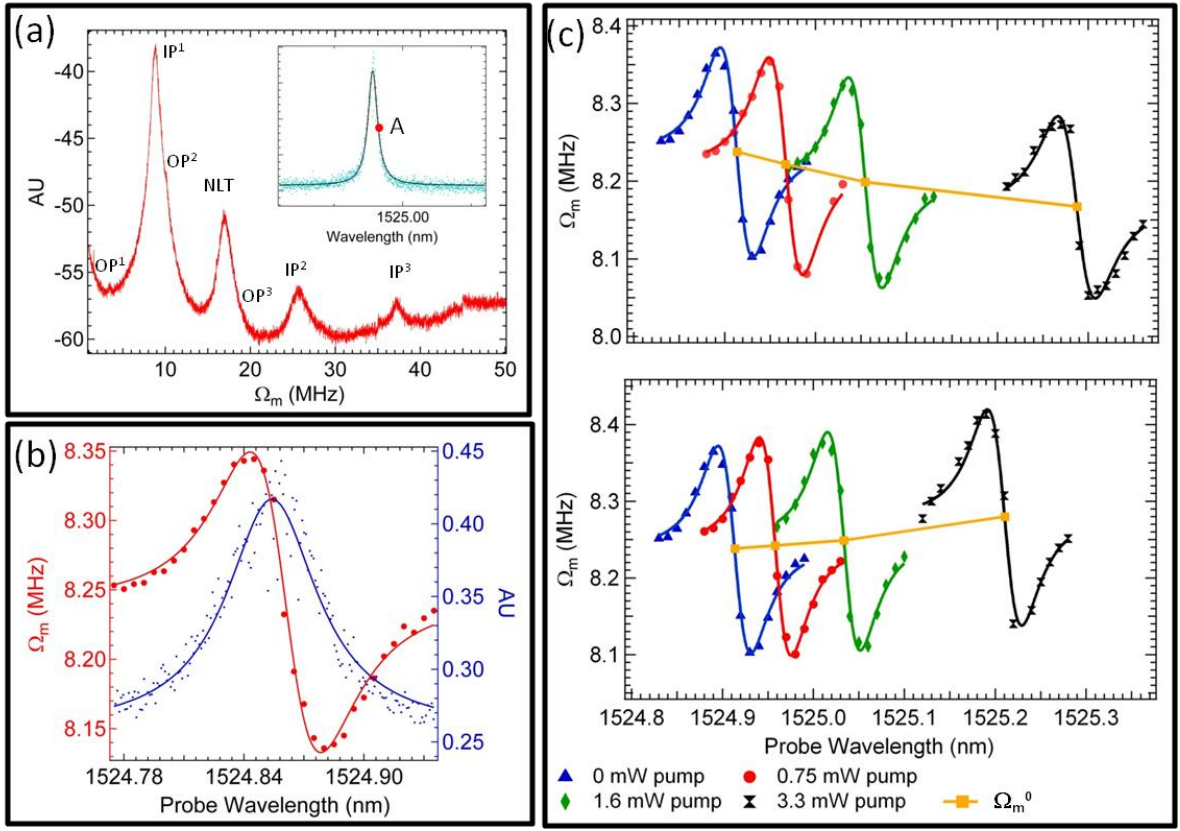

**Figure 4:** (a) Mechanical response spectrum showing the Brownian motion of the coupled PCNC. The inset shows the optical spectrum with operating point (pt A), corresponding to the maximum slope of the transmission resonance, indicated. The peaks are labeled according to their mechanical deflection which can be in-plane (IP) or out-of-plane (OP). The most efficient transduction is obtained for fundamental in-plane mode $IP^1$ (supplementary section 9.0). The peak at 16 MHz does not correspond to any mechanical mode, but is a Fourier component due to nonlinear transduction. This happens when Brownian motion of the beams is large enough to detune the probe laser across the lorentzian line shape of the cavity resonance (Supplementary section 9). In this case, the relation between optical transmission and gap is no longer linear around the operating point A. (b) Mechanical response of the fundamental in-plane mechanical mode – $IP^1$ - (red) for various detuning of the probe mode (blue) at zero *pump* power. This effect is due to self stiffening of the nanobeams due to large optical energy stored inside the cavities. (c) The shift in mechanical response shown in (b) for various *pump* power at two different *pump* detuning. The top graph is for a red detuned *pump* (1570.5 nm), while the bottom graph if for a blue detuned *pump* (1573.5 nm). The mechanical response moves towards lower frequencies for red detuning and towards higher frequencies for a blue detuned *pump*. The yellow curve shows the mechanical frequency of the probe without self-stiffening from the probe.

Next, we study the dynamical effects in our system. We characterized the Brownian motion of the two free standing beams, in ambient conditions, by detuning the probe to the maximum slope of the cavity resonance (pt. A in Fig.4a inset). Since the even mode of the coupled cavity system is extremely dispersive with respect to the gap between the beams, any motion of the beams is transduced onto the transmission of the cavity. This modulated transmission of the probe was analyzed using a spectrum analyzer and is shown in Fig. 4a. Various in-plane and out-of-plane vibrational modes can be identified in the spectrum. The fundamental in-plane mode which has the maximum $g_{om}$ shows up as a very strong peak at 8 MHz. The second and third order in-plane modes can also be well resolved in the spectrum. We note that the peak at 16 MHz does not correspond to any vibrational mode and is an artifact of nonlinear transduction (Section 9 in supplementary material). Fig. 4b plots the effect of probe detuning on the resonant frequency of the fundamental, mechanical, in-plane mode.

The blue curve shows the optical transmission spectrum of the even mode while the red curve plots the detuning of the mechanical resonance for various probe detuning. Since the device operates in the sideband unresolved limit ($\Omega_M \ll$ optical Q), the response is only due to optical stiffening[1] and is given by:

$$\Omega_M^2 = \Omega_{M0}^2 - \text{Re}(A_0 + B_0), \text{ where}$$

$$A_0 = -\frac{2\Delta_c'(g_{Om}^c)^2|a_c|^2}{\omega_c m_{eff} f_c f_c'}, \text{ and } B_0 = -\frac{2\Delta_p'(g_{Om}^p)^2|a_p|^2}{\omega_p m_{eff} f_p f_p'} \quad (2)$$

$$f_{c,p} \cong -i\Omega_{M0} - i\Delta_{c,p}' + \frac{\Gamma_{c,p}'}{2}$$

$$f_{c,p}' \cong -i\Omega_{M0} + i\Delta_{c,p}' + \frac{\Gamma_{c,p}'}{2}$$

where $\Omega_M$ is the mechanical Eigen frequency, $g_{om}$ is the optomechanical coupling constant, $m_{eff}$ is the effective mass of the mechanical mode, $\Delta_{c,p}'$ are the detuning of the input laser, $|a_{c,p}|^2$ are the energies stored in the respective modes and . $\Gamma_{c,p}'$ is the decay rate of the stored energy in the optical modes.

Apart from the self-stiffening effect due to the probe mode as seen in Fig. 4b, stiffening can also be induced by the *pum*p. This effect is studied by pumping at either side of the broad *pump* mode. As the power in the *pump* mode is increased the mechanical resonance of the device is detuned due to stiffening. The sign of this mechanical resonance detuning is dependent on the detuning of the *pump* laser with respect to the *pump* resonance (Fig. 4c). For a red-detuned *pump* (top) the beams are softened and the mechanical resonance frequency decreases, while blue detuning results in an increase of the mechanical frequency. The smooth curves show the fits using Eqn. 2 taking into account the presence of the *pump*. Therefore, our approach allows us to independently control, the optical and mechanical frequencies of our optomechanical filters. We note that in addition to attractive forces discussed above, repulsive force can also be obtained by exciting odd pump modes (Section 4 in supplementary material). This force, however, is much weaker due to small $g_{om}$ of the odd mode in our structure and does not result in observable effects.

In conclusion, we demonstrated an all-optical tunable filter employing coupled photonic crystal nanobeam cavities, where a considerable portion of the tuning is affected by mechanical reconfiguration. Specifically, we measured a total detuning of filter resonance of more than 18 line-widths out of which 20 % was due to opto-mechanical effects. Our scheme allows for a wide range of pump wavelength selection while overcoming the limitation of self-detuning of the pump and can be used to simultaneously control multiple devices. At the same time, independent control over the mechanical response was also achieved by varying the pump laser's wavelength. Our technique is a promising candidate for realization of reconfigurable and programmable optical devices, including filters and filter banks, routers, and modulators.

**Acknowledgements**

This work is supported in part by Defense Advanced Research Projects Agency (DARPA) under Contract No. N66001-09-1-2070-DOD and NSF CAREER grants. I.W.F. acknowledges support by NSF graduate research fellowship. Device fabrication was performed at the Center for Nanoscale Systems at Harvard and the Cornell Nanofabrication Facility.

# Supplementary Material
## 1. Device Fabrication and Characterization:

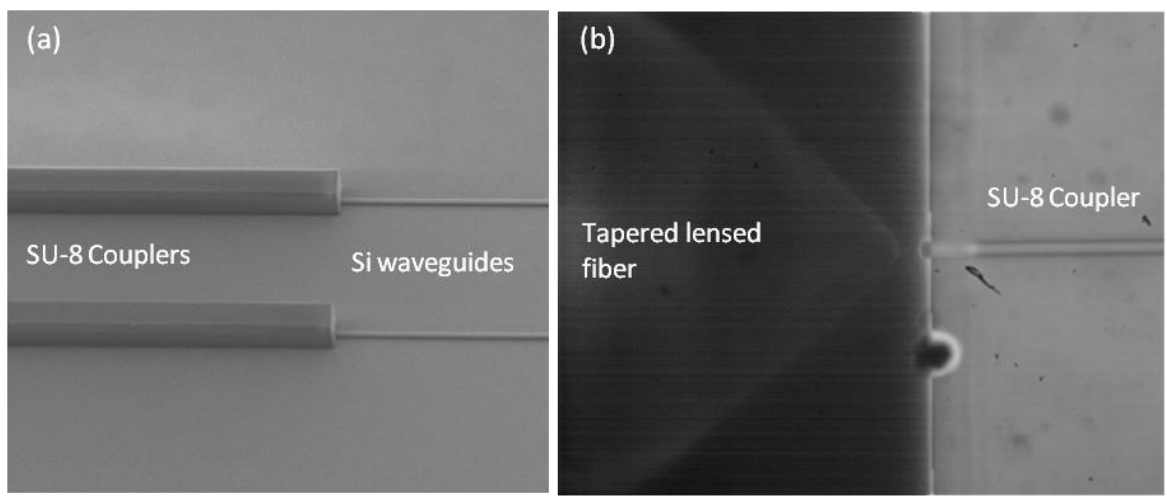

**Figure S5: (a) SEM image of the mode converter showing SU-8 coupler and Si waveguides (b) Optical micrograph of the cleaved facet of the silicon chip showing the tapered lensed fiber and the SU-8 coupler**

The devices were fabricated on a SOI substrate with a silicon device layer of 220 nm and a buried oxide layer of 3 μm. The patterns were transferred on to silicon device layer by using negative e-beam resist, XR-1451-002, with a 100 keV e-beam lithography tool. Proximity correction was necessary in order to achieve correct hole sizes for narrow waveguide gaps. The patterns were then etched into the Silicon using ICP-RIE with a Chlorine based chemistry. In the second fabrication step, SU-8 polymer coupling pads were fabricated on each end of the device to couple light on and off the chip. After this step, Shipley 1813S photoresist was spun on the sample. The area excluding the SU-8 pads was exposed using photolithography. The remaining ebeam resist covering the waveguides was removed by dipping the sample in 7:1 BOE for 1 minute. The sample was then cleaved through the SU8 pads and another photolithography step was carried out to open a small window in order to suspend the nanobeam photonic crystal cavities (PCNC). Finally, the cavities were under-etched using 7:1 BOE and then dried in a critical point dryer. The etch time was adjusted to achieve an etch depth of 2 μm.

Light was coupled using PM tapered lensed fiber (Oz optics, NA 0:4) into the SU-8 waveguide (refer Fig. S1a). An inverse taper geometry was used as a spot-size converter to efficiently transfer the light from SU-8 waveguides in to the silicon waveguides. This was necessary because SU-8 waveguides has significantly larger dimensions (3 x 3 μm) when compared to silicon waveguides (440 nm by 220 nm). At the output, a similar coupling method was used except the light was collected with a SM tapered lensed fiber (refer Fig. S1b) (SM tapered fibers have lower taper conversion loss). The propagation loss of the silicon waveguides and the maximum bend loss for 10 μm bend radius were measured to be 4.5 dB/cm and 1.0 dB, respectively. This was measured by imaging the light scattered from the waveguide from top with a sensitive IR camera. The SU8 waveguide propagation loss was measured with the same method and found to be 7 dB/cm. This is likely due to the excitation of higher order modes of the SU8 waveguides that are less confined and prone to higher losses. By measuring the insertion loss for the waveguide samples and using the measured propagation losses of the waveguides, we estimated the coupling loss associated with the spot-size converters to be 2 ± 0.5 dB. This was further confirmed by reversing the input signal direction. The power in the waveguide just before the cavity was estimated using the above measured values of coupling and propagation losses and the known device dimensions. The energy stored inside the cavity (W) is determined by taking into account the total Q ($Q_t$) and the waveguide coupling Q ($Q_c$) for the particular mode: W= 4*$Q_t^2 P_{in}/(\omega_0 Q_c)$.

The devices were characterized using tunable lasers (Agilent 81682, Santec TSL-510) and a sensitive InGaAs detector (EO systems, IGA-010-TE2-H). For mechanical measurements, a fast InGaAs detector (Newport 1811-FC) was used along with an external amplifier (Stanford Research Systems) before analyzing the signal using an electrical spectrum analyzer (Tektronix RSA3003). The pump laser signal was amplified using an L band 1 W EDFA (Manligth). The EDFA output was filtered for ASE emissions using a tunable filter (Agiltron). The pump and probe signals were multiplexed and de-multiplexed using a WDM (Microoptics Inc).

## 2. Thermal and Non-linear effects

We modeled our structure by using a two-mode, temporal, coupled-mode theory [1]:

$$\frac{da_c}{dt} = i\Delta'_c a_c - \frac{\Gamma'_c}{2} a_c + \sqrt{\Gamma_c^{wg}} s_c^{in} \qquad \text{S1}$$

$$\frac{da_p}{dt} = i\Delta'_p a_p - \frac{\Gamma'_p}{2} a_p + \sqrt{\Gamma_p^{wg}} s_p^{in} \qquad \text{S2}$$

$$\Gamma'_c = \Gamma_c^0 + \beta_c |a_c|^2 + \beta_{cp} |a_p|^2 + \alpha_c^{cc} N^{cc} + \alpha_c^{pp} N^{pp} + \alpha_c^{cp} N^{cp} \qquad \text{S3}$$

$$\Delta'_c = \omega_c^l - \omega_c^0 + \Delta_c^{TH} + \Delta_c^{FCD} + \Delta_c^{\chi^3} + \Delta_c^{Mech.} \qquad \text{S4}$$

$$\Gamma'_p = \Gamma_p^0 + \beta_p |a_p|^2 + \beta_{pc} |a_c|^2 + \alpha_p^{cc} N^{cc} + \alpha_p^{pp} N^{pp} + \alpha_p^{cp} N^{cp} \qquad \text{S5}$$

$$\Delta'_p = \omega_p^l - \omega_p^0 + \Delta_p^{TH} + \Delta_p^{FCD} + \Delta_p^{\chi^3} + \Delta_p^{Mech.} \qquad \text{S6}$$

where subscripts $c$ and $p$ refer to the high-Q probe mode and low-Q pump mode, respectively. $|a_{c,p}|^2$ are the energies stored in the respective modes, $|s_{c,p}^{in}|^2$ are equal to the input powers. $\Gamma'_{c,p}$ characterizes the decay rate of the stored energy in the optical modes, whereas $\Delta'_{c,p}$ are the detuning of the input laser frequency with respect to the resonant frequencies of the respective modes. We have included the following effects, which are relevant to silicon at telecommunication wavelengths: degenerate two-photon absorption ($\beta_{c,p} |a_{c,p}|^2$), non-degenerate two photon absorption ($\beta_{cp,pc} |a_{p,c}|^2$), free-carrier absorption ($\alpha_{c,p}^{cc} N^{cc} + \alpha_{c,p}^{pp} N^{pp} + \alpha_{c,p}^{cp} N^{cp}$), thermal dispersion $\Delta_{c,p}^{TH}$, free-carrier dispersion ($\Delta_{c,p}^{FCD}$), self-phase and cross-phase modulation ($\Delta_{c,p}^{\chi^3}$), and optomechanical detuning ($\Delta_{c,p}^{Mech.}$) in our model. $\Gamma_c'$ includes the cold cavity decay rates (waveguide coupling, scattering, and linear absorption), and non-linear absorption rates (degenerate ($\beta_c |a_c|^2$) and non-degenerate two-photon absorption ($\beta_{cp} |a_p|^2$), and free-carrier absorption rates ($\alpha_c^{cc} N^{cc} + \alpha_c^{pp} N^{pp} + \alpha_c^{cp} N^{cp}$) due to the free-carriers generated by two-photon absorption of the probe and pump mode photons). The non-linear absorption terms can be expressed in terms of effective mode parameters [2]. For instance, the modal degenerate two-photon absorption coefficient can be written as:

$$\beta_c = \frac{\Gamma_c^{TPA} \beta_{Si} c^2}{V_c^{TPA} n_g^2}$$

$$\Gamma_c^{TPA} = \frac{\int_{Si} \varepsilon^2(r) E_c^4(r) dv}{\int \varepsilon^2(r) E_c^4(r) dv} \qquad \text{S7}$$

$$V_c^{TPA} = \frac{\left[\int \varepsilon(r) E_c^2(r) dv\right]^2}{\int \varepsilon^2(r) E_c^4(r) dv}$$

where $\beta_{Si}$ is the bulk degenerate two-photon absorption coefficient. The modal non-degenerate two photon absorption coefficient can be written as:

$$\beta_{cp} = \frac{\Gamma_{cp}^{TPA} 2\beta_{Si} c^2}{V_{cp}^{TPA} n_g^2}$$

$$\Gamma_{cp}^{TPA} = \frac{\int_{Si} \varepsilon^2(r) E_c^2(r) E_p^2(r) dv}{\int \varepsilon^2(r) E_c^2(r) E_p^2(r) dv} \qquad \text{S8}$$

$$V_{cp}^{TPA} = \frac{\left[\int \varepsilon(r) E_c^2(r) dv\right]\left[\int \varepsilon(r) E_p^2(r) dv\right]}{\int \varepsilon^2(r) E_c^2(r) E_p^2(r) dv}$$

Note that the bulk non-degenerate two-photon absorption coefficient is twice the degenerate two-photon absorption coefficient [3]. In steady state, the free-carrier absorption rates can be expressed in terms of the stored energies and effective modal parameters. We assume a simplified model of carrier generation and diffusion such that the carriers generated by two-photon absorption of the pump and probe modes are treated independently. For instance, $\alpha_c^{cc} N^{cc}$, the free-carrier absorption rate of the probe mode due to the free-carriers generated by the two photon absorption of the probe photons, can be written as $\gamma_c^{cc} |a_c|^4$, where

$$\gamma_c^{cc} = \frac{\tau_{fc}\sigma_{Si} c \Gamma_c^{cc,FCA}}{n_g V_c^{cc,FCA}} \frac{\beta_{Si} c^2}{n_g^2 2\hbar\omega_c^0}$$

$$\Gamma_c^{cc,FCA} = \frac{\int_{Si} \varepsilon^3(r) E_c^6(r) dv}{\int \varepsilon^3(r) E_c^6(r) dv} \qquad \text{S9}$$

$$V_c^{cc,FCA} = \frac{\left[\int \varepsilon(r) E_c^2(r) dv\right]^3}{\int \varepsilon^3(r) E_c^6(r) dv}$$

Similarly, $\alpha_c^{pp} N^{pp}$, the free-carrier absorption rate of the probe mode due to the free-carriers generated by the two photon absorption of the pump photons, can be written as $\gamma_c^{pp}|a_p|^4$, where

$$\gamma_c^{pp} = \frac{\tau_{fc}\sigma_{Si} c \Gamma_c^{pp,FCA}}{n_g V_c^{pp,FCA}} \frac{\beta_{Si} c^2}{n_g^2 2\hbar\omega_p^0}$$

$$\Gamma_c^{pp,FCA} = \frac{\int_{Si} \varepsilon^3(r) E_c^2(r) E_p^4(r) dv}{\int \varepsilon^3(r) E_c^2(r) E_p^4(r) dv} \qquad \text{S10}$$

$$V_c^{pp,FCA} = \frac{\left[\int \varepsilon(r) E_c^2(r) dv\right]\left[\int \varepsilon(r) E_p^2(r) dv\right]^2}{\int \varepsilon^3(r) E_c^2(r) E_p^4(r) dv}$$

The detuning term is much more complicated and has a richer structure. The contribution due to the free-carrier dispersion can be decomposed into their respective free-carrier source terms: degenerate two photon absorption of the pump and probe modes, and the non-degenerate two-photon absorption:

$$\Delta_c^{FCD} = \Delta_c^{cc,FCD} + \Delta_c^{pp,FCD} + \Delta_c^{pc,FCD} \qquad \text{S11}$$

In steady state, we can express eqn. S11 as:

$$\Delta_c^{FCD} = g_c^{cc,FCD}|a_c|^4 + g_c^{pp,FCD}|a_p|^4 + g_c^{pc,FCD}|a_c|^2|a_p|^2 \qquad \text{S12}$$

where,

$$g_c^{cc,FCD} = -\frac{\omega_c^0 \zeta_{Si,e} \tau_{fc} \beta_{Si} c^2 \Gamma_c^{cc,FCD}}{2\hbar\omega_c n_{Si} n_g^2 V_c^{cc,FCD}}, \quad \frac{\Gamma_c^{cc,FCD}}{V_c^{cc,FCD}} = \frac{\Gamma_c^{cc,FCA}}{V_c^{cc,FCA}}$$

$$g_c^{pp,FCD} = -\frac{\omega_c^0 \zeta_{Si,e} \tau_{fc} \beta_{Si} c^2 \Gamma_c^{pp,FCD}}{2\hbar\omega_p n_{Si} n_g^2 V_c^{pp,FCD}}, \quad \frac{\Gamma_c^{pp,FCD}}{V_c^{pp,FCD}} = \frac{\Gamma_c^{pp,FCA}}{V_c^{pp,FCA}} \qquad \text{S13}$$

$$g_c^{cp,FCD} = -\frac{\omega_c^0 \zeta_{Si,e} \tau_{fc} 2\beta_{Si} c^2 \Gamma_c^{cp,FCD}}{\hbar(\omega_p + \omega_c) n_{Si} n_g^2 V_c^{cp,FCD}}, \quad \frac{\Gamma_c^{cp,FCD}}{V_c^{cp,FCD}} = \frac{\Gamma_c^{cp,FCA}}{V_c^{cp,FCA}}$$

The above equations explain FCD contribution from electrons. Note that there is a similar contribution from holes with ($\bar{N}^{0.8}$) dependence.

The contribution due to optomechanical detuning can be written as:

$$\Delta_c^{Mech.} = -g_{om}^c x$$

$$\Delta_c^{Mech.} = \frac{(g_{om}^c)^2 |a_c|^2}{\Omega_M^2 m_{eff} \omega_c^0} + \frac{g_{om}^c g_{om}^p |a_p|^2}{\Omega_M^2 m_{eff} \omega_p^0}$$

S14

Where x is the displacement amplitude of the optically bright mechanical mode, $g_{om}$ is the opto-mechanical coupling constant, $\Omega_M$ is the mechanical eigen-frequency, and $m_{eff}$ is the effective mass of the mechanical mode. The choice of the displacement amplitude *x*, is somewhat arbitrary. We have adopted the convention introduced in ref. S3. The Optomechanical coupling constant is calculated from perturbation theory of shifting boundaries [2-3]. For the device under test, the optomechanical coupling coefficient ($g_{om}$) was found to be 96 GHz/nm for the even mode. Note that we have ignored the beat terms between the pump and probe modes. This is justified within the rotating wave approximation because the frequency separation between the pump and probe modes are much larger than the mechanical frequency. We have also verified that the above equation describes the displacement of the nano-beams due to the optical forces correctly. The forces due to an excitation of the pump or probe mode were calculated and applied as a boundary load in FEM simulations. The displacement obtained from the FEM simulations (refer Fig. S2) was compared to the above model. We have found that both the methods agree within 2-5%, which is well within the numerical accuracy of the calculations.

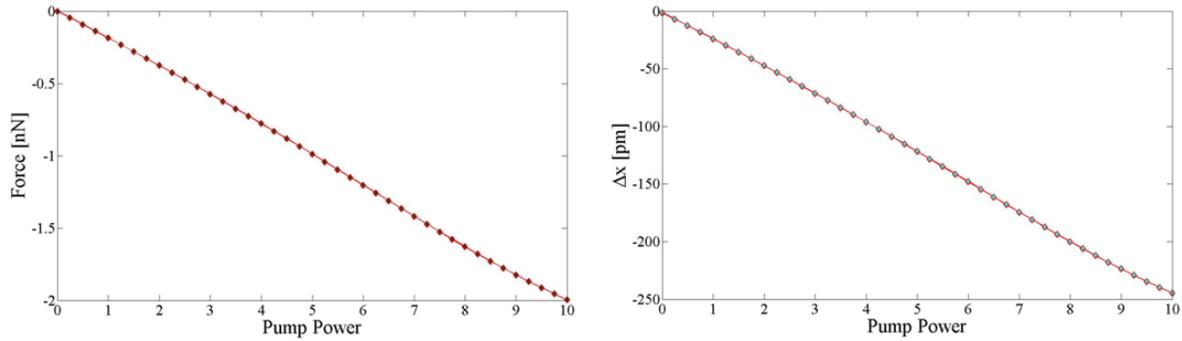

**Figure S6: Estimated force and displacement for different pump powers (power inside the waveguide just before the cavity (mW)).**

We have also considered both self-phase modulation and cross-phase modulation due to the Kerr effect:

$$\Delta_c^{Kerr} = \chi^c |a_c|^2 + \chi^{cp} |a_p|^2$$

S15

where first term is due to the self-phase modulation and the second term is due to cross-phase modulation. And

$$\Delta_c^c = \frac{\omega_c^0 \Gamma_c^{Kerr} n_{2,Si} c}{n_{Si} V_c^{Kerr} n_g}, \quad \Gamma_c^{Kerr} = \Gamma_c^{TPA}, \quad V_c^{Kerr} = V_c^{TPA}$$

$$\Delta_c^{cp} = \frac{\omega_c^0 \Gamma_{cp}^{Kerr} 2 n_{2,Si} c}{n_{Si} V_{cp}^{Kerr} n_g}, \quad \Gamma_{cp}^{Kerr} = \Gamma_{cp}^{TPA}, \quad V_{cp}^{Kerr} = V_{cp}^{TPA}$$

S16

Finally, the thermal dispersion can be written as:

$$\Delta_c^{TH} = \frac{\omega_c^0 \Gamma_c^{th}}{n_{Si}} \frac{dn_{Si}}{dT} \left( T_c + T_p \right)$$

$$\Gamma_c^{th} = \frac{\int_{Si} \varepsilon(r) E_c^2(r) dv}{\int \varepsilon(r) E_c^2(r) dv}$$

$$T_{c,p} = R_{c,p}^{th} P_{abs}(|a_{c,p}|^2)$$

S17

where, $T_c$ ($T_p$) is the temperature change due to the absorption of the probe photons (pump photons). $R_{c,p}^{th}$ is the thermal resistance of the structure when the heat is generated due to photons absorbed from the respective probe or pump modes. $P_{abs}(|a_{c,p}|^2)$ is the total power absorbed from the respective modes. The thermal resistances for both modes were estimated from FEM simulations. The thermal conductivity in thin and unpatterned SOI samples was measured by several groups, and it was found that the thermal conductivity strongly depends on the thickness of the sample [S4-S6]. A thermal conductivity of 40 W/mK has been reported for 115 nm thick silicon samples[5], which is ~4 times smaller than that of bulk Si. We expect an even smaller effective thermal conductivity in our devices due to the patterning. A thermal conductivity as low as 15 has been reported for poly-silicon samples with 210 nm grains, which is of a similar length scale as our feature dimensions. In our device, the mean hole-to-hole separation was around 170 nm and the mean distance between the edge of the nano-beam and the holes was 125 nm. A thermal conductivity of 20 W/mK was used in FEM simulations. This value was further confirmed by comparing FEM simulations with thermal time constant measurements. The thermal time constant was measured by a pump-probe method, where the pump was modulated via an electro-optic modulator. A thermal time constant of 80 kHz was measured for our devices.

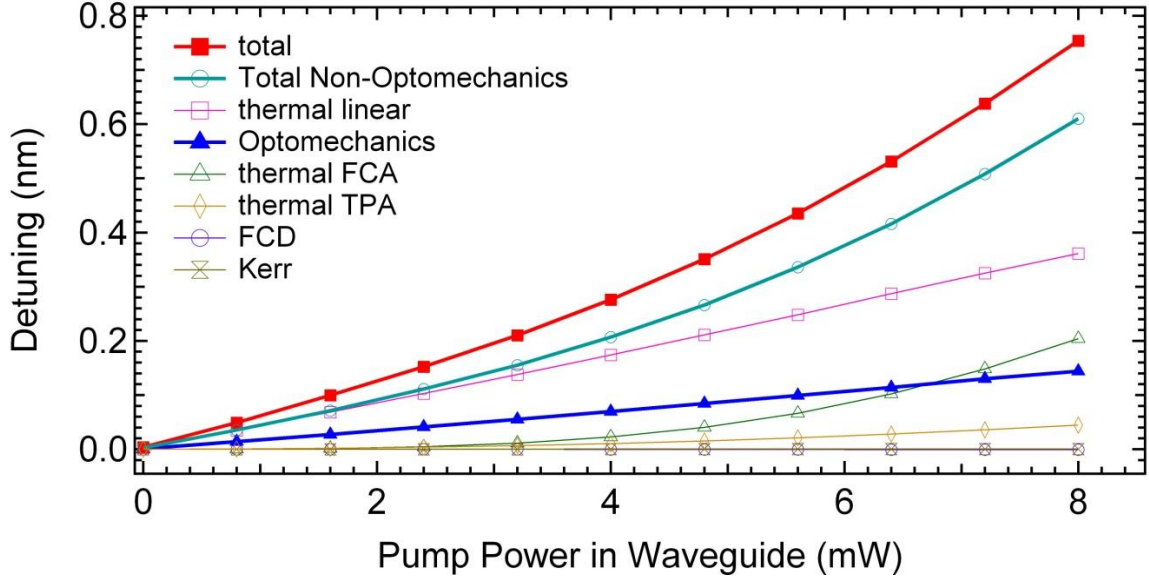

**Figure S7: Estimated detuning of probe for different pump powers (Power in the waveguide just before photonic crystal cavity).**

The transmission properties and the detuning of the probe mode are determined from the steady-state solutions of eqns. S1 and S2. The list of parameters used in our calculations is summarized in table S1 and the estimate for different detuning is plotted in Fig. S3.

## 2.1 Dynamics

The dynamical response of the system can be determined from the following system of coupled equations:

$$\frac{da_c}{dt} = i\Delta'_c a_c - \frac{\Gamma'_c}{2} a_c + \sqrt{\Gamma_c^{wg}}\, s_c^{in} \qquad \text{S18}$$

$$\frac{da_p}{dt} = i\Delta'_p a_p - \frac{\Gamma'_p}{2} a_p + \sqrt{\Gamma_p^{wg}}\, s_p^{in} \qquad \text{S19}$$

$$\frac{dT}{dt} = -\frac{T}{\tau_{th}} + c_{th} P_{abs}(|a_p|^2) \qquad \text{S20}$$

$$\frac{dN}{dt} = -\frac{N}{\tau_{fc}} + \frac{\beta_p |a_p|^4}{2\hbar\omega_p} \qquad \text{S21}$$

$$\frac{d^2 x}{dt^2} + \Gamma_M \frac{dx}{dt} + \Omega_M^2 x = -\frac{g_{OM}^c |a_c|^2}{m_{eff}\omega_c} - \frac{g_{OM}^p |a_p|^2}{m_{eff}\omega_p} + F_L \qquad \text{S22}$$

Equation S22 is written within rotating-wave approximation [S9]. Since a very weak probe beam (0.1 µW) was used in our experiments, we have excluded the non-linear effects induced by the probe mode from this model. These equations can be solved perturbatively using the following ansatz [S7]:

$$\begin{aligned}
a_c &= \overline{a_c} + \delta a_c \\
a_p &= \overline{a_p} + \delta a_p \\
T &= \overline{T} + \delta T \\
N &= \overline{N} + \delta N \\
x &= \overline{x} + \delta x
\end{aligned} \qquad \text{S23}$$

where ($\overline{a_c}$, $\overline{a_p}$, $\overline{T}$, $\overline{N}$, $\overline{x}$) are the solutions of the steady-state equations. In the perturbative limit, the general procedure to solve these equations is that ansatz S23 is substituted in eqns. S18-S22, only the linear dynamical terms are kept, and finally ($\delta a_c$, $\delta a_p$) are expressed in terms of $\delta x$ and substituted in eqn.S22 [S10]. In the limit of weak probe input and low-Q pump the solution can be simplified and the contribution of each mode can be added separately to the effective mechanical frequency:

$$\Omega_M^2 = \Omega_{M0}^2 - \mathrm{Re}(A_0 + B_0), \text{ where}$$

$$A_0 = -\frac{2\Delta_c' (g_{Om}^c)^2 |\overline{a_c}|^2}{\omega_c m_{eff} f_c f_c'}, \text{ and } B_0 = -\frac{2\Delta_p' (g_{Om}^p)^2 |\overline{a_p}|^2}{\omega_p m_{eff} f_p f_p'}$$

$$f_{c,p} \cong -i\Omega_{M0} - i\Delta_{c,p}' + \frac{\Gamma_{c,p}'}{2}$$

$$f_{c,p}' \cong -i\Omega_{M0} + i\Delta_{c,p}' + \frac{\Gamma_{c,p}'}{2} \qquad \text{S24}$$

$$\Delta_{c,p}' = \Delta_{c,p}^0 + g_{c,p}^{TH}\overline{T} + g_{c,p}^{FCD}\overline{N} + \chi^{c,p}\left|\overline{a_{c,p}}\right|^2 + \chi^{cp,pc}\left|\overline{a_{p,c}}\right|^2 - g_{OM}^{c,p}\overline{x}$$

$$\Gamma_{c,p}' = \Gamma_{c,p}^0 + \alpha_{c,p}\overline{N} + \beta_{c,p}\left|\overline{a_{c,p}}\right|^2 + \beta_{cp,pc}\left|\overline{a_{p,c}}\right|^2$$

These equations and the effect of the strong pump can be understood in the following way. First, at low probe input powers without an excitation in the pump mode, we observe optical spring effect. Once the pump mode is excited at a constant wavelength, first it detunes the probe mode, and reduces the energy stored in probe mode because of non-linear absorption (non-degenerate two photon absorption, and free-carrier absorption). Moreover, the pump mode contributes a constant term to the effective mechanical frequency (the pump wavelength is kept constant). Finally, as the pump power is increased there is a self-detuning effect, i.e. the pump mode detunes itself.

**Table 1: List of all the simulation parameters**

| Quantity | Value | Unit | Ref |
|---|---|---|---|
| $\beta_{Si}$ | 7.9e-12 | m/W | S8 |
| $V_c^{TPA}$ | 5.81e-19 | $m^3$ | |
| $V_p^{TPA}$ | 4.96e-19 | $m^3$ | |
| $V_{cp,pc}^{TPA}$ | 3.18e-18 | $m^3$ | |
| $V_c^{cc,FCA}$ | 2e-37 | $m^3$ | |
| $V_c^{pp,FCA}$ | 5e-36 | $m^3$ | |
| $V_c^{cp,FCA}$ | 1.3e-36 | $m^3$ | |
| $V_p^{cc,FCA}$ | 1.3e-36 | $m^3$ | |
| $V_p^{pp,FCA}$ | 2.37e-36 | $m^3$ | |
| $V_p^{cp,FCA}$ | 5e-36 | $m^3$ | |
| $\Delta'_{c,p}$ | 90 | GHz | |
| $\sigma_{Si}$ | 1.45e-21 | $m^2$ | S8 |
| $n_{2,Si}$ | 5e-18 | $m^2/W$ | S8 |
| $\zeta_{Si,e}$ | 8.8e-28 | $m^3$ | S1,9 |
| $\zeta_{Si,h}$ | 1.45e-29 | $m^3$ | S1,9 |
| $dn_{Si}/dT$ | 1.86e-4 | $K^{-1}$ | |
| $\tau_{fc}$ | 2200e-12 | s | |
| $m_{eff}$ | 3.059e-15 | kg | |
| $g_{Om}^c$ | 96e9 | GHz/nm | |
| $R_{c,p}^{th}$ | 1.17e6 | K/W | |
| $Q_{abs}$ | 1e6 | | |
| $Q_c$ | 25000 | | |
| $Q_p$ | 360 | | |

## 3. Thermal Tuning and Calibration of the Even and Odd Modes:

Silicon has a relatively large thermo-optic coefficient, $dn_{Si}/dT = 1.86 \times 10^{-4} K^{-1}$. As a result, any change in temperature modifies the resonance frequency of the cavity, Eqn. S17. This temperature detuning is slightly different for each mode because of different mode overlap factors with silicon, Eqn. S17. Hence, it is important to characterize the tuning due to temperature for each mode, especially for the mode used for on-chip temperature calibration. This was determined experimentally by measuring the resonance frequency of both the even and odd modes as a function of temperature. The sample was placed on a thermo-electric heater and the temperature on the sample surface was measured. The ratio of wavelength detuning with temperature of the two modes was determined from the slope of the temperature versus detuning curves. The experimentally determined value of tuning ratio of the even mode to the odd mode was found to be 0.93. This value was used in all the calculations.

## 4. Excitation using an un-balanced Mach-Zehnder Interferometer:

We considered only TE polarized modes throughout this work. Our devices support both highly localized cavity modes with large Q-factors and extended waveguide-like modes with low Q-factors. Due to the symmetry of the structure with respect to mirror plane in the middle of the gap (perpendicular to the substrate), both localized and extended modes have either even or odd symmetry. The high-Q modes were considered as probe/filter and low-Q modes were used to pump and actuate the devices. We aimed to tune the resonance frequency of the filter modes by using optomechanical actuation when the devices were excited with a low-Q pump mode. Fig S4 shows the theoretical transmission spectrum, for both modes with even and odd symmetry, of an actual fabricated device used to obtain the data shown in Fig 3 and 4. The dimensions were determined with scanning-electron microscopy (SEM) measurements, and the measured dimensions were used in finite-element (FEM) simulations. The bottom plots show the electric fields associated with high-Q probe and low-Q pump modes. The probe modes are localized while the pump modes are extended and resemble waveguide like modes.

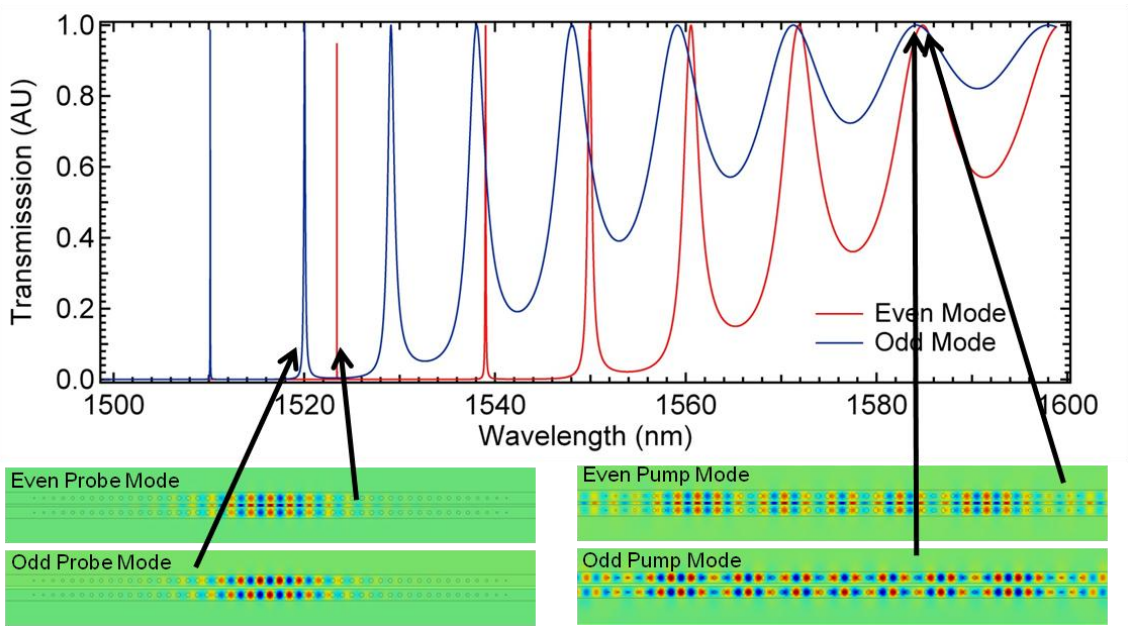

**Figure S8: Simulation of the optical transmission of an actual device (dimensions measured using SEM) when excited with the even (red) and odd (blue) symmetry mode. The pump and probe electric fields for the even and odd mode are also shown in the bottom. Probe modes are localized while the pump modes resemble waveguide modes.**

In order to excite the even and odd modes simultaneously, we incorporated an unbalanced Mach-Zhender Interferometer (MZI) at the input of the device (200 μm path difference). An unbalanced MZI generates a phase difference between its output arms. This phase difference depends on the wavelength and the difference between the lengths of the MZI arms. When the phase difference is 0, even mode is excited, and when the phase is π the odd mode is excited. For phase difference in the range between 0 and π both even and odd modes are excited. The signal at the output (after the cavities) from each of the waveguides was collected without combining them. This allowed us to excite and collect both even and odd modes of the structure. For experimental results shown in Fig. 2, we decided to use a balanced MZI to split the power equally in the two waveguides instead of redesigning a 3 dB coupler.

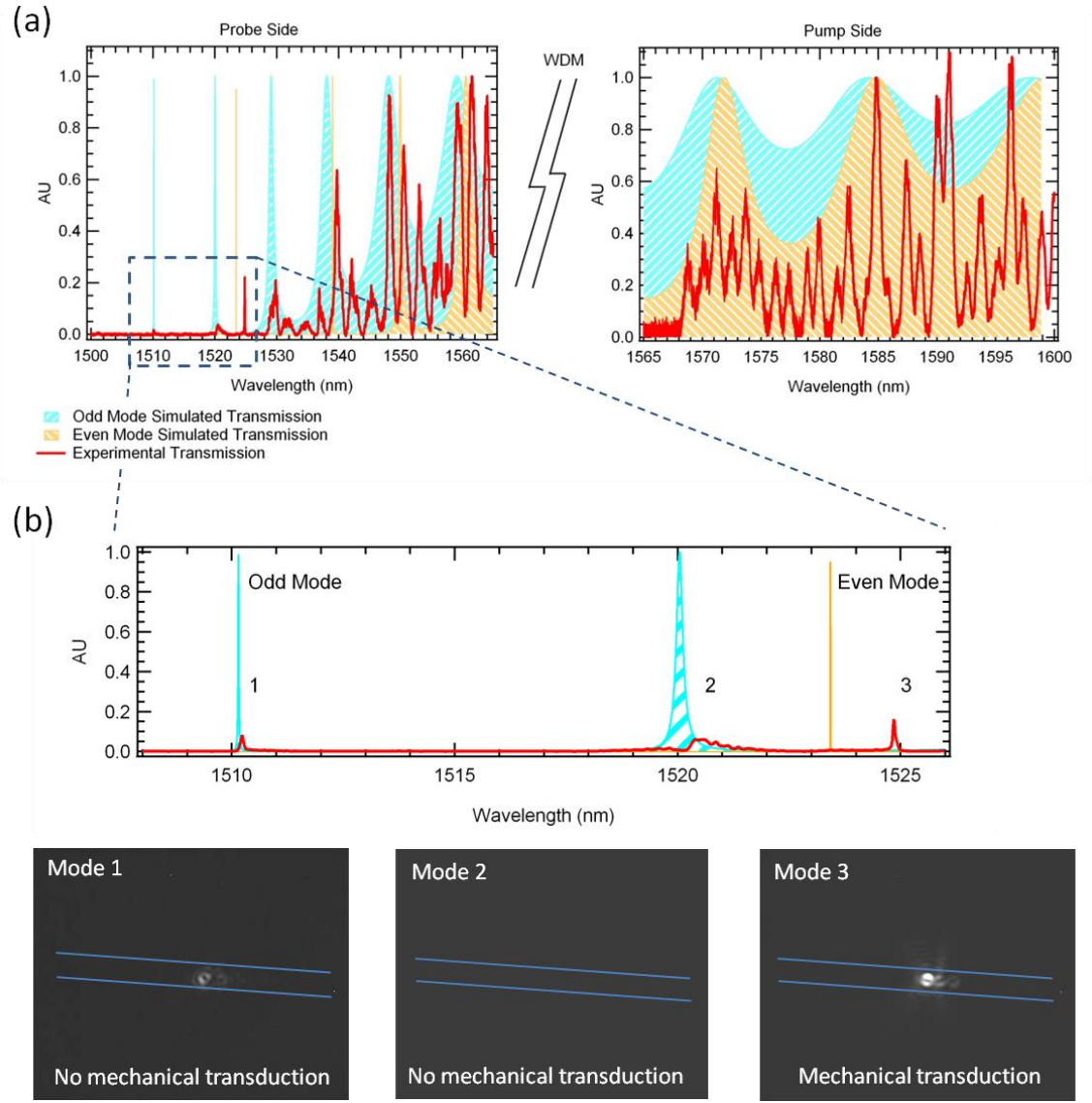

**Figure S9: (a) Experimental transmission spectrum through a device with an unbalanced MZI at the input. Superimposed on it is the simulated transmission shown in Fig. S5. The actual transmission is the even/odd MZI fringes enveloped by even/odd transmission of the cavity. The two spectra correspond to each side of the WDM filter arm. It should be pointed out that the y-axis in both the cases is different. (b) Zoomed area showing the even and odd modes. The broad resonance around 1520.5 nm is possibly the higher order odd mode. We confirmed the even and odd modes by imaging them from top using a sensitive IR camera. The images show the localized nature of the modes. The two blue lines indicate the boundary of the two beams. In order to confirm the symmetry of the localized modes, we used them to transduce mechanical motion of the beams due to Brownian motion. The even mode is expected to have good transduction properties (due to large $g_{om}$) while the odd mode is expected to have poor transduction property. In addition, odd mode is expected to be found at shorter wavelengths, based on the FDTD simulations. Therefore, by taking advantage of mode imaging, motion transduction, and spectral characteristics of the mode, we were able to unambiguously identify even and odd modes of the structure, and found them to be in good agreement with theoretical predictions.**

The measured transmission spectrum through a device with an input MZI is shown in Fig S5a along with the simulated transmission. The spectrum consists of the even/odd MZI fringes superimposed on the even/odd transmission spectrum of the cavity. The fundamental odd and even modes were found to be at 1510.1 nm and 1524.48 nm, respectively (Fig. S5b). To confirm this, cavities were excited at these three wavelengths (Fig. S5b, top panel) and imaged using a sensitive IR camera (Fig. S5b, bottom panel). Our results indicate that the modes at 1510.1 nm and 1524.48 nm are localized in the cavity region. Furthermore, mechanical transduction was observed only with the mode at 1524.48 nm. These observations, along with the fact that the odd mode was expected to be at lower wavelength as compared to the even mode, allowed us to associate the mode at 1524.48 nm with the even high-Q fundamental mode, and the 1510.1 nm mode with the high-Q odd mode.

## 4.1 Force due to Odd mode:

As can be seen from Fig. 1b, the dispersion for the odd mode is extremely small. As a result the $g_{om}$ for the odd mode is about two orders of magnitude smaller when compared to the even mode. This holds true even for the odd low-Q waveguide like modes used as the pump (checked using numerical modeling). Therefore, repulsive force resulting from the odd pump modes is orders of magnitude smaller than the attractive force induced by the even mode. It is interesting to note that the dispersion of the localized odd cavity mode (probe) switches sign at about 110 nm of separation between nanobeams. Below this gap, the force generated by the odd probe mode is attractive while it is repulsive for larger gaps. However, it must be noted that even this force is extremely small (especially for larger gaps) and is not detectable in the current setup with the available pump power levels: we calculated the expected tuning due to the localized odd cavity mode to be about 20 times smaller in our devices with 70 nm separation.

# 5. Tuning by using a high Q cavity mode:

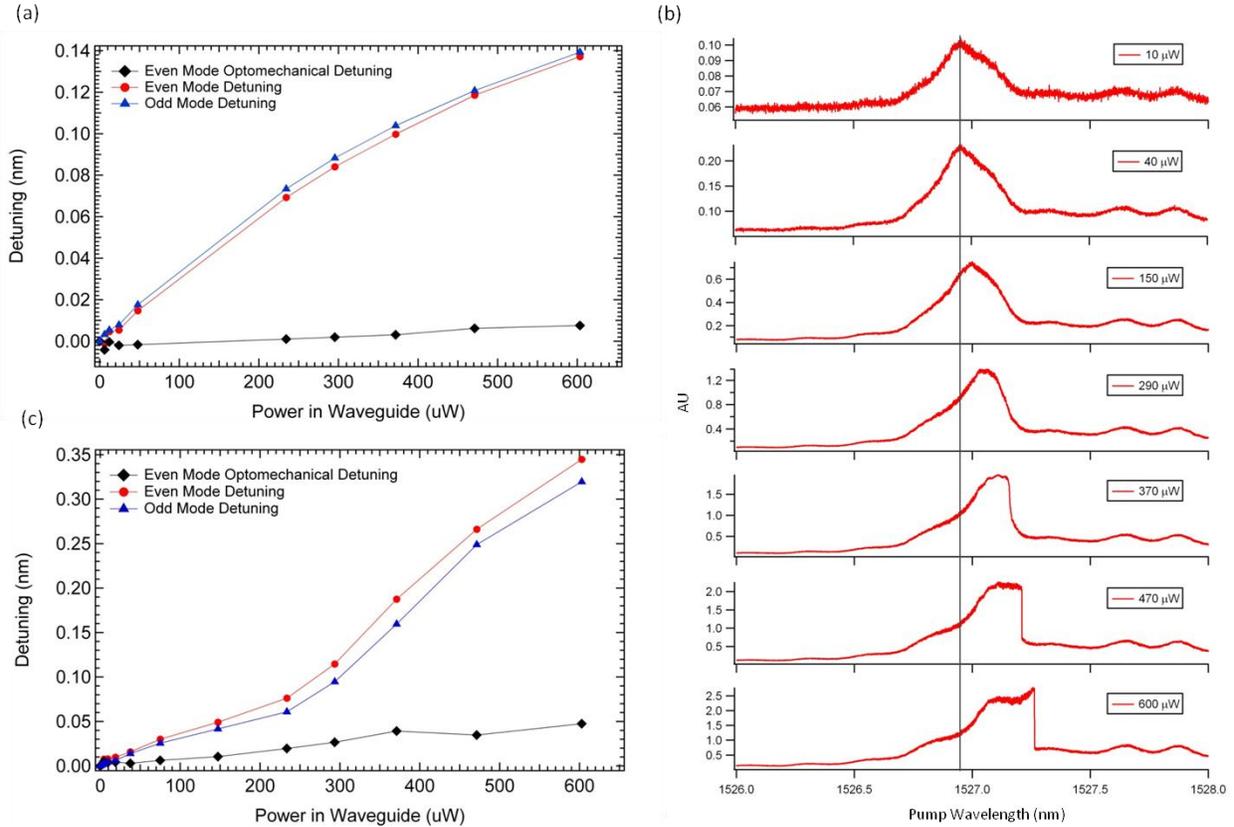

**Figure S10:** High Q cavity mode pumping (a) even and odd mode detuning for various pump powers without pump mode tracking and locking. (b) Detuning of the pump mode for various pump powers. Self detuning of the pump mode is observed with increase in pump power. Bistable operation due to optical nonlinearity can be seen for power levels of 370 μW and higher (c) even and odd mode detuning for various pump powers and with pump mode tracking and locking. The detuning achieved in this case is 2.5 times larger for the same power levels as compared to (a).

We also used high-Q modes in order to actuate the beams and tune the resonant wavelength of the filters. The results of these experiments confirm the problems associated with using a high Q mode as pump, and show that broadband operation is not possible without wavelength tracking and locking. These experiments were conducted on a different device but with comparable gap between the beams. The particular pump mode had a Q of ~ 5500. Fig. S6a shows the detuning of the even (1515.4 nm) and odd (1508.2nm) modes and the estimated optomechanical detuning as a function of input pump power (power inside the waveguide before the cavities). The pump wavelength was fixed at the maximum transmission point (at low pump powers) of the pump mode at 1526.96 nm. This wavelength corresponds to the 2$^{nd}$ even resonance of the structure. Comparing this with the detuning shown in Fig. 3b, we see that it requires about 3 times less power to achieve the same detuning when the device is excited with this particular high-Q mode. However, the tuning saturates at large input powers due to self-detuning of the pump mode. Fig S6b shows the transmission spectrum around the pump mode for various input pump powers. We can clearly see that the pump mode self-detunes towards longer wavelengths with increasing powers. In addition, strong thermo-optic nonlinearity can be observed due to two-photon absorption of Si. Finally, we employed a wavelength tracking and locking technique. The spectrum was measured at each power level, and the laser wavelength was locked to the new resonant wavelength. With this method, 2.5 times larger tuning (Fig. S6c) is achieved at the same input power levels when compared to Fig. S6a (no tracking and locking). Hence, in order to achieve larger tunings the use of wavelength tracking and locking methods are required. The self-detuning has a larger effect at higher input powers, which is necessary for large tunings. The effects of bistability in our devices can be observed from Fig S6b for pump powers above 370 μW. Bistability becomes a large problem for high-Q devices as the power required for bistable operation scales with $1/Q^2$. These results clearly illustrate that unless a track and lock method is employed, it is not a feasible choice to use high-Q modes for achieving large tunings in silicon devices.

## 6.     Reduced thermal effects by pulsed excitation:

Thermo-optical effects are significant in silicon nano-photonic devices at telecommunication wavelengths. We observe this in our experiments when we compare the detuning due to thermo-optical effects and opto-mechanical effects (Fig. 3 in the main text). One way of reducing the relative contribution of thermo-optical effects in our devices is to improve $g_{om}$ by using nanobeams with smaller gaps. But, this is challenging with our current fabrication capabilities. Another method, discussed in this section, is to actuate the beams by using optical pulses with repetition rates faster than the thermal response time. We expect to observe three major operation regimes when pulsed excitation is used, defined by the thermal and mechanical response times (we only consider the fundamental mechanical mode here). In our device, the thermal cutoff frequency, which determines the thermal response time, is below 100 kHz while the resonance of fundamental mechanical mode is around 8 MHz with a $Q_{mechanical}$ = 17. In the first operation regime, the modulation frequency is below the thermal cutoff frequency and as a result both thermo-optical and optomechanical effects can follow the modulation. In the second region, the modulation frequency is larger than the thermal cutoff frequency but smaller than mechanical resonance. Therefore, the device settles at a thermal steady-state point while mechanical motion can follow the modulation. Therefore, tuning due to optomechanics becomes dominant in this modulation range. Finally, in the third region the modulation frequency is larger than both thermal and mechanical cutoff frequencies, and neither effect can keep up with the modulation, and the response of the system converges to zero.

In order to confirm these predictions, the pump was modulated by using an electro-optical modulator (a modulation depth of ~ 40 % was used, which was limited by the modulator). A 50 % duty cycle square wave was applied using a signal generator. The response of the even and odd modes was characterized with a tunable laser and a sensitive but slow detector (1 kHz bandwidth). The slow detector was particularly useful in capturing the detuning of the resonances for modulation frequencies higher than 1 kHz: for modulation frequencies higher than 1 kHz, the optical signal on the detector changed faster than the response time of the detector (or the cavity resonance was detuned at a rate much faster than the response time of the detector), and the detector averaged out all fluctuations and generated a cleaner signal. We also note that the laser was scanned at a speed of 5 pm/ms, and therefore the detector produced an averaged output of the optical signal for every 5 pm of cavity detuning.

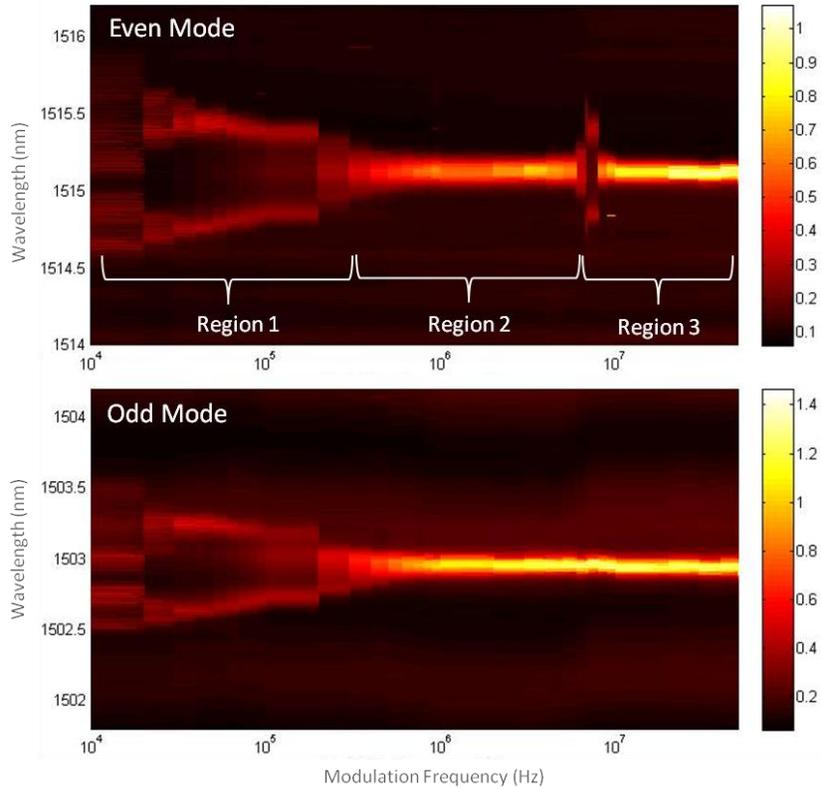

**Figure S7: Detuning of the even and odd modes for various pump modulation frequencies at constant pump power. Three regions of operation, defined by the thermal cut-off and mechanical cut-off frequencies can be identified. Region 1 denotes the regime where the thermo-optic and optomechanical effects can follow the modulation signal. Region 2 denotes the regime above the thermal cut-off frequency, where the thermal effects settle to a steady state value and only the optomechanical effects are prominent. Region 3 denotes the regime above the mechanical cut-off frequency, where both the thermo-optic and optomechanical effects are too slow to follow the modulation signal. The high detuning in the even mode at the mechanical frequency (~ 8 MHz) is due to enhancement of the motion because of mechanical Q but is absent in the case of the odd mode due to its low $g_{om}$.**

The three regions of operation mentioned above can clearly be seen in Fig. S7. Region 1 represents the operation regime where detuning due to thermal and optomechanical effects are prominent. This regime extends from dc operation to thermal cut-off frequency. Here cavity resonance is modulated over wide wavelength range, due to combination of thermal and mechanical effects, and it appears smeared-out by the slow detector. Region 2 is the regime where the thermal effects settle to a steady state value and only optomechanical detuning remain prominent. This regime extends from thermal cut-off to mechanical cut-off frequency. This is precisely the regime where opto-mechanics is the dominant tuning mechanism. The wavelength "smearing" is smaller in this region than in Region 1 which is consistent with our observations that thermal effects cause roughly four times more tuning than mechanical effects in the case of DC actuation. Also, as expected there is no smearing in the odd-mode resonance in this regime since its $g_{om}$ is very low. Finally, in Region 3, which is above the mechanical cut-off frequency, neither thermo-optical nor mechanical effects can follow the modulation. At mechanical frequency, the motion is amplified by the mechanical Q resulting in a large detuning of the even mode while this is not observed for the odd mode. Again, this is consistent with the much higher $g_{om}$ of the even mode.

# 7. Optical spring effect and back-action (cooling/amplification of the mechanical mode):

Fig. S8 illustrates the effect of optical mode on the frequency, amplitude and linewidth of the fundamental mechanical mode as a function of laser wavelength. The change in mechanical frequency of the fundamental mode is due to optical spring effect. We did not observe any parametric instability for the power levels used in this work. Furthermore, thermal instabilities were also absent. No appreciable cooling (broadening of the linewidth) and amplification (narrowing of the linewidth) was observed for red/blue detuning of the probe mode.

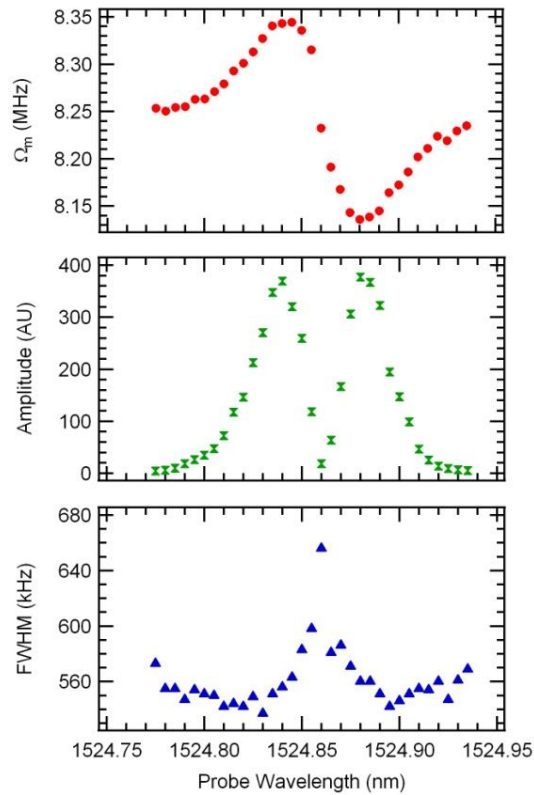

**Figure S8: Fundamental mechanical mode frequency, amplitude and linewidth for various optical detunings of the probe mode**

## 8. Measurement of the Optomechanical Coupling Constant:

In order to experimentally determine $g_{om}$ and verify the accuracy of our theoretical model, we used the frequency noise calibration technique with a direct detection method [S14]. A calibration signal of known modulation depth was generated by modulating the probe using a phase modulator at a frequency close to the mechanical resonance as shown in Fig. S9. The peak amplitude of the calibration signal and the mechanical resonance was noted. A correction to the measured mechanical resonance peak as reported in [S14] was applied before estimating the $g_{om}$. Unfortunately, the device with 70 nm gap, used in Fig. 3 and 4, that has theoretically calculated $g_{om}$ of 96 GHz/nm was damaged before $g_{om}$ measurements could be performed. Therefore, a different device was used to carry out the experiment and verify the $g_{om}$ values predicted using our theoretical model. The gap between the nanobeams for the device under test was measured to be 111.8 nm. Theoretical value of $g_{om}$ for this gap was $38 \pm 3$ GHz/nm (error associated due to measuring dimensions from SEM image). The measured value obtained using the noise calibration approach was found to be 35 GHz/nm. Such excellent agreement between the experimental and simulation value confirms the accuracy of our model and justifies the use of the theoretically calculated value of $g_{om}$ in our modeling.

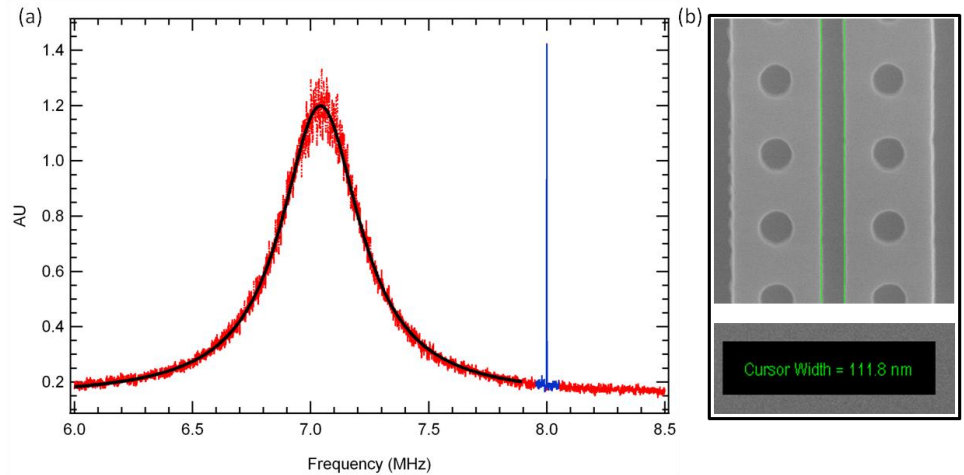

Figure S9: (a) Mechanical spectrum showing the fundamental mechanical mode along with the calibration signal. The black line is the Lorentzian fit for the mechanical mode while the blue line is a Gaussian fit for the calibration signal as mentioned in Ref. S14. (b) SEM image of the device used for $g_{om}$ measurement

## 9. Non-Linear Transduction of Mechanical modes:

The mechanical spectrum of the suspended nanobeams can be analyzed by probing the cavity resonance at the maximum slope point A as shown in the Fig. S10. The various vibrational modes labeled in Fig. S10 were identified using FEM simulations. The displacement of the beams has been exaggerated to depict motion.

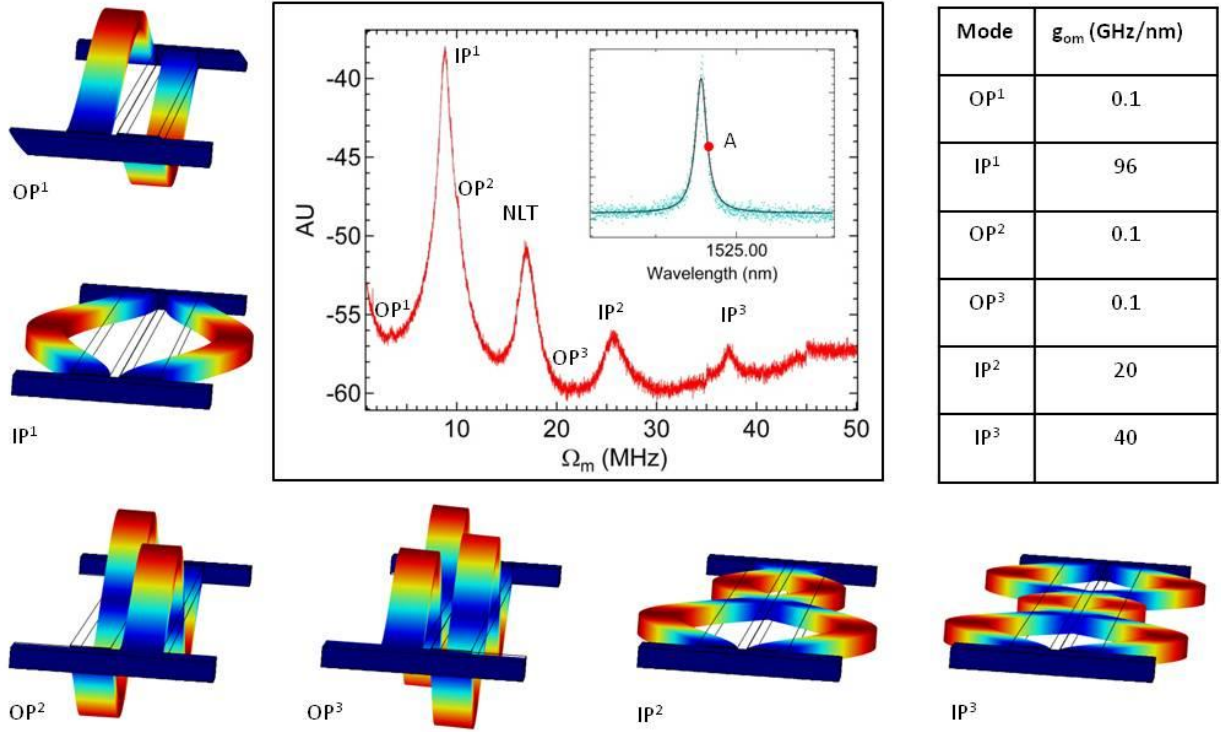

**Figure S10: Mechanical spectrum.** The spectrum was obtained by probing the cavity resonance at the maximum slope point as shown in the inset. Various vibrational modes were identified using FEM simulations. The nanobeam displacements have been exaggerated to show the vibrational motion of the beams. The table summarizes the $g_{om}$ of various modes for a device with 70 nm gap between the nanobeams.

The $g_{om}$ (calculated) for various modes have been summarized in table shown in Fig. S11. Comparing the simulated results with the observed mechanical spectrum reveals a strong peak at 16 MHz which does not correspond to any of the theoretically expected modes.

We investigated this further by conducting similar experiment on a different sample with similar cavity dimensions (except with a slightly longer suspended length and larger separation ~ 112 nm, $g_{om}$ 38 GHz/nm). This measurement was done on different sample since the one used to obtain the data in the main manuscript was unfortunately destroyed. The top plot in Fig. S11 shows the mechanical spectrum transduced using a high-Q (~35000) cavity optical mode. Using the equipartition theorem, we estimate that the Brownian motion of the beams gives rise to about 38 pm of motion. With the known $g_{om}$ (38 GHz/nm) this translates approximately 12 pm of change in wavelength. The HWHM of the probe resonance is 21 pm (Q ~ 35000). This change in wavelength due to motion results in nonlinear transduction of the signal.

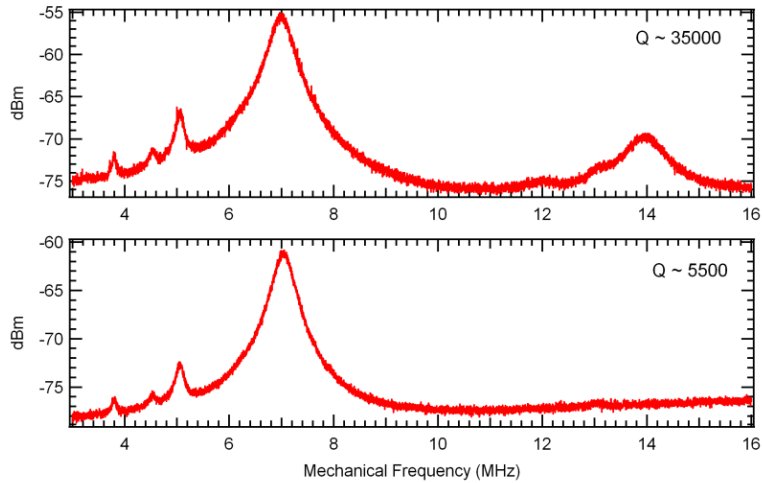

**Figure S11: Mechanical spectrum transduced using a high Q and low Q optical mode. The 14 MHz peak in the top plot is a Fourier component of the fundamental mode at 7 MHz. This was due to nonlinear transduction, which was confirmed by analyzing the signal transduced using a low Q optical mode (lower plot). The $g_{om}$ of both the optical modes were comparable (38 GHz/nm and 35 GHz/nm).**

The nonlinear transduction effect is illustrated in Fig. S12, where the inset shows the deviation of the measured signal from a linear response due to lorentzian lineshape of the optical mode. This gives rise to the peak at 14 MHz which is a non-linear harmonic of the fundamental mode at 7 MHz (Fig. S11). The lower graph in Fig. S11 shows mechanical spectrum transduced using a low Q optical mode (Q ~ 5500, $g_{om}$ 35 GHz/nm) where the peak at 14 MHz is missing confirming the nonlinear transduction in the earlier case.

Similarly, the resonance at 16 MHz in Fig. S10 is a harmonic of the fundamental mode due to nonlinear transduction ($g_{om}$ 96 GHz/nm, Q ~ 25,000). The effect in this sample is more pronounced due to small nanobeam gap resulting in higher $g_{om}$.

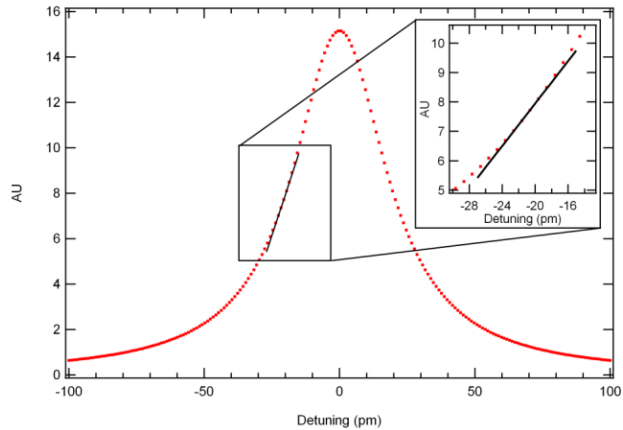

**Figure S12: Lorentzian lineshape for an optical Q of 35000. Brownian motion of the nanobeams results in 12 pm of detuning around the operating point, which is plotted in black. Inset shows the deviation of the signal from a linear response for the above detuning.**

# Supplementary References: